\documentstyle[12pt]{article}
\begin{document}

\title{Towards possibility of self-maintained vacuum traversible
wormhole}
\author{V. Khatsymovsky \\
 {\em Budker Institute of Nuclear Physics} \\ {\em Novosibirsk,
 630090,
 Russia} \\ {\em E-mail address: khatsym@inp.nsk.su}}
\date{}
\maketitle
\begin{abstract}
We calculate renormalised vacuum expectation values of
electromagnetic stress-energy tensor in the static
spherically-symmetrical wormhole topology. We find that for
metric tensor sufficiently slow varied with distance violation
of the averaged weak energy condition takes place irrespectively
of the detailed form of metric. This is a necessary condition
for the electromagnetic vacuum to be able to support the
wormhole geometry.
\end{abstract}
\newpage
{\bf 1.Introduction.} The possibility of existence of static
spherically-symmetrical \\ traversible wormhole as
topology-nontrivial solution to the Einstein equations have been
first studied by Morris and Thorne in 1988 \cite{MT}. They have
found that the material which threads the wormhole should violate
weak energy condition at the throat of the wormhole, that is,
radial pressure should exceed the density. Moreover, Morris,
Thorne and Yurtsever \cite{MTY} have pointed out that averaged
weak energy condition (i.e. that integrated over the radial
direction) should also be violated. They have also suggested
Casimir vacuum between conducting spherical plates as an example
of such the material which is able to support the wormhole.

As the next step, it would be natural to check the possibility
of existence of self-maintained vacuum wormhole threaded by
vacuum fields in the wormhole geometry in the absence of any
conductors. For that one must find vacuum expectation
value of the stress-energy tensor as functional of geometry and
solve Einstein equations with this tensor as a source. Both
parts of this strategy (explicit finding stress-energy tensor
and solving the equations) are very difficult problems.
Calculation of vacuum expectations of the stress-energy tensor
for different fields in the curved background has been considered
in a number of works. Recently functional dependence of
stress-energy on the metric is known only in the form of some
anzats approximately valid for some class of metrics. For
example, Page, Brown and Ottewill \cite{P}, \cite{BO},
\cite{BOP} have suggested approximate expression for
stress-energy of the conformal massless scalar, spinor and
vector fields in the Einstein spacetimes (where
$R_{\mu\nu}=\Lambda g_{\mu\nu}$,
$\Lambda ={\rm const}$). In some cases this approximation can
rather accurately reproduce numerical results \cite{HC},
\cite{H} while in another cases it may disagree with precise
values \cite{JO}. Frolov and Zel'nikov \cite {FZ} have suggested
another anzats for the stress-energy of conformally invariant
massless fields in static spacetimes.

In the absence of exact expression  for the stress-energy as
functional of metric it seems to us that the most useful for our
purposes would be an expansion over some parameter (which can be
small, at least in principle). Consider quasiclassical expansion
which is now expansion over derivatives of metric (over radial
coordinate). In \cite{Kh} the renormalised stress-energy tensor
of electromagnetic vacuum has been calculated with the help of
covariant geodesic point separation method of regularization
\cite{Christ}. It has been found to violate weak energy
condition at the wormhole throat in the first nonvanishing
order in the expansion over the derivatives of metric. This is a
necessary condition of existence of self-maintained wormhole
solution. Important is that this violation takes place
irrespectively of the detailed form of metric. As for other
fields, there may be those which oppose the maintenance of the
certain wormhole metrics. Anderson, Hiscock and Samuel
\cite{AHS1}, \cite{AHS2} have developed a method for numerical
calculation of the stress-energy tensor of a quantised scalar
field with arbitrary curvature coupling and mass in a general
static spherically symmetrical spacetime. Using this method,
Anderson, Hiscock and Taylor \cite{AHT} have found for five test
wormhole geometries that both minimally or conformally coupled
massive scalar field vacuum cannot maintain these geometries
since it does not violate weak energy condition at the throat.
Thus massive scalar field will hardly be of interest in the
self-maintained wormhole aspect.

In the given note we continue studying the electromagnetic
vacuum case and find that in the first nonvanishing order in the
expansion over derivatives also averaged weak energy condition
is violated in the wormhole topology and this violation takes
place irrespectively of the detailed form of metric. Moreover,
the Einstein equations formally admit some wormhole type
solution, namely, that corresponding to the infinitely long
wormhole (however, there is no authomatical smallness in the
higher orders of expansion over derivatives in this case, and
corresponding approximation well may be unacceptable).

\bigskip
{\bf 2.Calculation.} The notations for the metric functions
$r(\rho)$, $\Phi(\rho)$ are read from the following expression
for the line element:

\begin{equation}
ds^2=r^2(\rho)(d\theta^2+\sin^2{\!\theta}\,d\phi^2)+d\rho^2
-\exp{(2\Phi)}dt^2.
\end{equation}
We split the point in the radial distance $\rho$ and consider
the values of interest as bilocals at $\rho$, $\tilde{\rho}$
such that 

\begin{equation}
\tilde{\rho}-\rho=\epsilon\rightarrow 0.
\end{equation}
We introduce also the variable $z$ via

\begin{equation}
dz=\exp{(-\Phi)}d\rho.
\end{equation}
Local values taken at $\tilde{\rho}$, $\rho$ are denoted by
tilded and untilded letters respectively, as, e.g.,
$\tilde{\Phi}$ and $\Phi$. The "physical" components are defined
as

\begin{equation}
A_{\hat{\mu}\hat{\nu}...\hat{\lambda}}\stackrel{\rm
def}{=}|g_{\mu\mu}g_{\nu\nu}...g_{\lambda\lambda}|^{-1/2}A_{\mu\nu...
\lambda}
\end{equation}
for any tensor $A$ (these are notations of ref.\cite{MT}).

Standard summation over electromagnetic modes leads to the
following expressions for the vacuum expectation values of the
split-regularized stress-energy tensor \cite{Kh}:

\begin{eqnarray}
\label{T-mu-nu}
4\pi\left (\matrix{ T_{\hat{t}\hat{t}}\cr
T_{\hat{\rho}\hat{\rho}}\cr T_{\hat{\theta}\hat{\theta}}\cr }
\right)^{\rm reg}=\sum_{l=1}^{\infty}{\int_{0}^{\infty}{{2l+1
\over\pi}dt}}\left\{\left[{\exp{(\Phi -\tilde{\Phi})}\over r^3
\tilde{r}}\left (\matrix{ 1\cr 1\cr 0\cr } \right )+{1\over
 r^2\tilde{r}^2}\left (\matrix{ 1\cr -1\cr 1\cr }\right )\right]
\cdot l(l+1)\right. \nonumber\\ \left. +\left (
\matrix{ 1\cr 1\cr 0\cr }\right ){\exp{(-\Phi-\tilde{\Phi})}
\over r\tilde{r}}{\partial
\over\partial z}\left ({\partial\over\partial\tilde{z}}-
{\partial\over\partial z}\right )
\right\}
G(t,l;z,\tilde{z})
\end{eqnarray}
(the vacuum expectations are denoted just as components
themselves: this will not lead to any confusion). The
$G(t,l;z,\tilde{z})$ is half of the sum of "electric" and
"magnetic" Green functions,

\begin{equation}
G=(G^E+G^M)/2,
\end{equation}
obeying the equation

\begin{equation}
\left[-{\partial^2\over\partial z^2}+t^2+l(l+1){\exp{2\Phi}
\over r^2}\right]G^{E,M}(t,l;z,\tilde{z})=\delta (z-\tilde{z})
\end{equation}
with the following boundary conditions on the conducting
spherical plates:

\begin{eqnarray}
G^E|_{z\in\Gamma}=0,\\
\left. {\partial G^M\over\partial z}\right| _{z\in\Gamma}=0.
\end{eqnarray}
The surface $\Gamma$ can be introduced as some auxiliary
infra-red regulator and then shifted to infinity (this is a way
to give a sence to summation over continuum set of energy levels
in the infinite space). Then it proves that

\begin{equation}
G^E=G^M=G.
\end{equation}

Solution to the nonuniform equation can be expressed in terms of
the solutions to the corresponding uniform equation $\Psi_+$ and
$\Psi_-$ which fall off at $z\rightarrow +\infty$ and
$z\rightarrow -\infty$, respectively. In turn, one of these, say
$\Psi_-$, can be expressed in terms of another one, $\Psi_+$. As
a result, 

\begin{equation}
G(z,\tilde{z})=\Psi_+(z)\Psi_+(\tilde{z})
\int_{-\infty}^{z}{dz\over \Psi^2_+(z)}
\end{equation}
at $\tilde{z}>z$. Let us denote $\Psi_+=\exp{(-\Omega)}$, write
$dz=d\Omega/\Omega^\prime$ and integrate by parts. Repeatedly
performing this procedure gives

\begin{equation}
\label{G(z,z')}
G(z,\tilde{z})=\exp{\left[-\int_{z}^{\tilde{z}}{\Omega^\prime (y)
dy}\right]}\sum_{k=0}^{\infty}{\left[-{1\over 2\Omega^\prime (z)}
{d\over dz}\right]^k{1\over 2\Omega^\prime (z)}}
\end{equation}
where $\Omega$ obeys the equation

\begin{equation}
\Omega^{\prime 2}-\Omega^{\prime\prime}=\omega^2(z)
\equiv t^2+l(l+1){\exp{(2\Phi)\over r^2}}.
\end{equation}
The solution to this equation is well-known from quantum
mechanics as quasiclassical series which up to the second order
reads:

\begin{equation}
\Omega^\prime=\omega +{\omega^\prime\over 2\omega}+
{\omega^{\prime\prime}\over 4\omega^2}-{3\over 8}
{\omega^{\prime 2}\over\omega^3}.
\end{equation}
Substitute this into (\ref{G(z,z')}). Integration over $dt$ and
summation over $l$ are finite due to the exponential factor
$\exp{(-\omega\Delta z)}$. Therefore the divergence at

\begin{equation}
\Delta z\equiv\tilde{z}-z\rightarrow 0
\end{equation}
arises in a few first terms in quasiclassical expansion.
Maximal divergence is $\Delta z^{-4}$, therefore the factors up
to $O(\Delta z^4)$ at these divergences also contribute into
finite part of the stress-energy. The resulting part of the
Green function of interest tuns out to be

\begin{eqnarray}
2G(z,\tilde{z})=\exp{(-\omega\Delta z)}\left\{{1\over\omega}-
{1\over 8}{(\omega^2)^{\prime\prime}\over\omega^5}+{5\over 32}
{(\omega^2)^{\prime 2}\over\omega^7}\right.\nonumber\\
+\Delta z\left [-{1\over 4}{(\omega^2)^\prime\over\omega^3}-
{1\over 8}{(\omega^2)^{\prime\prime}\over\omega^4}+{5\over 32}
{(\omega^2)^{\prime 2}\over\omega^6}\right ]\nonumber\\
+\Delta z^2\left [-{1\over 4}{(\omega^2)^\prime\over\omega^2}-
{1\over 8}{(\omega^2)^{\prime\prime}\over\omega^3}+{5\over 32}
{(\omega^2)^{\prime 2}\over\omega^5}\right ]\nonumber\\
\left.+\Delta z^3\left [-{1\over 12}{(\omega^2)^{\prime\prime}
\over\omega^2}+{5\over 48}{(\omega^2)^{\prime 2}\over\omega^4}
\right ]+\Delta z^4{1\over 32}{(\omega^2)^{\prime 2}\over
\omega^3}\right\}.
\end{eqnarray}
Upon substituting this into the expressions for the
stress-energy (\ref{T-mu-nu}) the terms like

\begin{equation}
\sum_{l=1}^{\infty}{\left(l+{1\over 2}\right)[l(l+1)]^m\int
_{0}^{\infty}{dt{\exp{\left [-\Delta z\sqrt{t^2+l(l+1)V(z)}
\,\right ]}\over [t^2+l(l+1)V(z)]^{n/2}}}}
\end{equation}
arise. Here

\begin{equation}
V\equiv{\exp{(2\Phi)}\over r^2}.
\end{equation}
Introduce instead of $t$ a new integration variable

\begin{equation}
q=\Delta z\sqrt{{t^2\over l(l+1)}+V(z)}
\end{equation}
and rewrite these terms in the form

\begin{equation}
\Delta z^{n-1}\int_{\Delta z\sqrt{V}}^{\infty}{{dq\over
 q^{(n-1)}\sqrt{q^2-\Delta z^2V}}\left (-{d\over dq}\right )
^{2m+1-n}{f(q)\over q^2}}
\end{equation}
where

\begin{equation}
f(q)\equiv q^2\sum_{l=1}^{\infty}{\left (l+{1\over 2}\right )
\exp{\left [-q\sqrt{l(l+1)}\right ]}}.
\end{equation}
The $f(q)$ is regular at $q\geq 0$. It turns out that
coefficients in $T_{\mu\nu}^{\rm reg}$ at $\Delta z^{-n}$, $n=0,
1,2,3,4$ and at $\ln{(L/\Delta z)}$, $L={\rm const}$, in
the considered quasiclassical orders can be expressed in terms
of $f^{(m)}(0)$, $m=0,1,2,3,4$. In turn, the latter are
elementary calculable by taking into account a few first terms in
the expansion of $\sqrt{l(l+1)}$ in the last formula over
decreasing powers of $l+1/2$ \cite{Kh}.

The result of calculations made according to the above scheme
gives for the two basic integrosums entering (\ref{T-mu-nu}):

\begin{eqnarray}
\sum_{l=1}^{\infty}{\int_{0}^{\infty}{\left (l+{1\over 2}
\right )dt\,l(l+1)G}}&=&{2\over\Delta z^4}{1\over V^2}-{2\over
\Delta z^3}{V^\prime\over V^3}+{1\over\Delta z^2}\left ({7
\over 4}{V^{\prime 2}\over V^4}-{5\over 6}{V^{\prime\prime}
\over V^3}\right )\nonumber\\
&&\hspace{-0.5cm}-{1\over 60}\left (\ln{{L\over \Delta z
\sqrt{V}}}-{1\over 2}\right )+{1\over 72}{V^{\prime\prime}
\over V^2}-{1\over 72}{V^{\prime 2}\over V^3},\\
\sum_{l=1}^{\infty}{\int_{0}^{\infty}{\left (l+{1\over 2}
\right )dt{\partial\over\partial z}\left ({\partial\over\partial
\tilde{z}}-{\partial\over\partial z}\right )G}}&=&-{6\over
\Delta z^4}{1\over V}+{4\over\Delta z^3}{V^\prime\over V^2}
+{1\over\Delta z^2}\left ({1\over 3}-{31\over 12}{V^{\prime 2}
\over V^3}\right.\nonumber\\
&&\hspace{-1cm}\left.+{4\over 3}{V^{\prime\prime}\over V^2}
\right )+{1\over 60}\left (\ln{{L\over \Delta z\sqrt{V}}}-1
\right )V-{1\over 144}{V^{\prime 2}\over V^2}.
\end{eqnarray}
Here the derivatives of $V$ are those over $\rho$ and numerical
constant under the logarithm sign is defined according to

\begin{equation}
\ln{L}={13\over 12}+\ln{M}-{5\over 2}\int_{0}^{\infty}
{{f^{\imath {\rm v}}(q)-f^{\imath {\rm v}}(0)\theta(M-q)
\over q}dq}
\end{equation}
where $f^{\imath {\rm v}}(0)=-2/5$ and $M\geq 0$ is arbitrary.

When substituting this into (\ref{T-mu-nu}) one should also
expand $\tilde{r}$, $\tilde{\Phi}$ at $z$ over $\Delta z$ and
express $\Delta z$ in terms of $\Delta\rho =\epsilon$. The
result of calculations is

\begin{eqnarray}
8\pi^2r^4T_{\hat{t}\hat{t}}^{\rm reg}&\hspace{-3mm}=&
\hspace{-3mm}-8{r^4\over\epsilon^4}
+{4\over 3}{r^2\over\epsilon^2}\left (1-r^{\prime 2}+2r^{\prime
\prime}r\right )-{4\over 3}{r\over\epsilon}r^{\prime}
-{1\over 15}\ln{{Lr\over\epsilon}}+{13\over 9}r^{\prime 2}
-{8\over 9}r^{\prime\prime}r,\\
8\pi^2r^4T_{\hat{\rho}\hat{\rho}}^{\rm reg}&\hspace{-3mm}=&
\hspace{-3mm}-24{r^4\over
\epsilon^4}+{4\over 3}{r^2\over\epsilon^2}\left [1+r^2\left (
2\Phi^{\prime\prime}+2\Phi^{\prime 2}-2\Phi^{\prime}{r^{\prime}
\over r}-{r^{\prime 2}\over r^2}+4{r^{\prime\prime}\over r}
\right )\right ]-{4\over 3}{r\over\epsilon}r^{\prime}\nonumber\\
&&\hspace{-3mm}+{1\over 15}\left (\ln{{Lr\over\epsilon}}
-{1\over 2}\right )
+{1\over 9}r^2\left (-2\Phi^{\prime\prime}-2\Phi^{\prime 2}
+2\Phi^{\prime}{r^{\prime}\over r}+11{r^{\prime 2}\over r^2}
-6{r^{\prime\prime}\over r}\right ),\\
8\pi^2r^4T_{\hat{\theta}\hat{\theta}}^{\rm reg}&\hspace{-3mm}=&
\hspace{-3mm}8{r^4\over
\epsilon^4}+{4\over 3}{r^2\over\epsilon^2}\left (-\Phi^{\prime
\prime}-\Phi^{\prime 2}+\Phi^{\prime}{r^{\prime}\over r}
-{r^{\prime\prime}\over r}\right )-{1\over 15}\left (
\ln{{Lr\over\epsilon}}-{1\over 2}\right )\nonumber\\&&
\hspace{-3mm}+{1\over 9}r^2\left (\Phi^{\prime\prime}
+\Phi^{\prime 2}
-\Phi^{\prime}{r^{\prime}\over r}+{r^{\prime 2}\over r^2}
-{r^{\prime\prime}\over r}\right ).
\end{eqnarray}

The divergences at $\epsilon\rightarrow 0$ are reduced to
renormalisation of the cosmological constant, Einstein gravity
constant and coefficient at the Weyl tensor squared in the
effective gravity action \cite{deWitt}. Thus we should equate
these coefficients to their experimental values and subtract
from $T_{\mu\nu}^{\rm reg}$ the $T_{\mu\nu}^{\rm div}$
corresponding to the divergent part of the effective action and
derived by Christensen \cite{Christ}. Calculation according to
the Christensen's formula for the electromagnetic field in our
geometry gives:

\begin{eqnarray}
8\pi^2r^4T_{\hat{t}\hat{t}}^{\rm div}&\hspace{-3mm}=&
\hspace{-3mm}-8{r^4\over\epsilon^4}
+{4\over 3}{r^2\over\epsilon^2}\left (1-r^{\prime 2}+2r^{\prime
\prime}r\right )-{4\over 3}{r\over\epsilon}r^{\prime}
-{1\over 15}\ln{{\Lambda\over\epsilon}}+r^{\prime 2}
-{1\over 3}r^{\prime\prime}r,\\
8\pi^2r^4T_{\hat{\rho}\hat{\rho}}^{\rm div}&\hspace{-3mm}=&
\hspace{-3mm}-24{r^4\over
\epsilon^4}+{4\over 3}{r^2\over\epsilon^2}\left [1+r^2\left (
2\Phi^{\prime\prime}+2\Phi^{\prime 2}-2\Phi^{\prime}{r^{\prime}
\over r}-{r^{\prime 2}\over r^2}+4{r^{\prime\prime}\over r}
\right )\right ]\nonumber\\
&&\hspace{-3mm}+{1\over 15}\left (\ln{{\Lambda\over\epsilon}}
-{1\over 2}\right )
+{1\over 9}r^2\left (-2\Phi^{\prime\prime}-2\Phi^{\prime 2}
-3\Phi^{\prime}{r^{\prime}\over r}+12{r^{\prime 2}\over r^2}
-6{r^{\prime\prime}\over r}\right ),\\
8\pi^2r^4T_{\hat{\theta}\hat{\theta}}^{\rm div}&\hspace{-3mm}=&
\hspace{-3mm}8{r^4\over
\epsilon^4}+{4\over 3}{r^4\over\epsilon^2}\left (-\Phi^{\prime
\prime}-\Phi^{\prime 2}+\Phi^{\prime}{r^{\prime}\over r}
-{r^{\prime\prime}\over r}\right )-{1\over 15}
\ln{{\Lambda\over\epsilon}}\nonumber\\&&
\hspace{-3mm}-{1\over 6}r^2\left (\Phi^{\prime\prime}
+\Phi^{\prime 2}\right ).
\end{eqnarray}
The coefficient at the Weyl term is the only one which is both
UV and IR divergent, and $\Lambda$ is IR cut-off. We include
both these divergences into the renormalisation of the
coefficient at the Weyl term. The Christensen's procedure
includes also forming half of the sum of the components
$T_{\mu\nu}$ corresponding to point separations $\epsilon =
\pm |\epsilon |$ (above formulas are given for $\epsilon >0$),
thereby only even powers of $|\epsilon |$ are left (and also
$|\epsilon|$ rather then $\epsilon$ enters the logarithm).
Subtracting $T_{\mu\nu}^{\rm div}$ from $T_{\mu\nu}^{\rm reg}$
we finally obtain

\begin{eqnarray}
\label{T-ren}
8\pi^2r^4T_{\hat{t}\hat{t}}^{\rm ren}&\hspace{-3mm}=&
\hspace{-3mm}-{1\over 15}\ln{{Lr\over\Lambda}}
+{4\over 9}r^{\prime 2}-{5\over 9}r^{\prime\prime}r,\\
8\pi^2r^4T_{\hat{\rho}\hat{\rho}}^{\rm ren}&\hspace{-3mm}=&
\hspace{-3mm}+{1\over 15}\ln{{Lr\over\Lambda}}
+{5\over 9}\Phi^{\prime}r^{\prime}r-{1\over 9}r^{\prime 2},\\
8\pi^2r^4T_{\hat{\theta}\hat{\theta}}^{\rm ren}&\hspace{-3mm}=&
\hspace{-3mm}-{1\over 15}\ln{{Lr\over\Lambda}}+{1\over 30}
+{5\over 18}r^2\left (\Phi^{\prime\prime}
+\Phi^{\prime 2}\right )
\nonumber\\&&\hspace{-3mm}
-{1\over 9}\Phi^{\prime}r^{\prime}r
+{1\over 9}r^{\prime 2}-{1\over 9}r^{\prime\prime}r.
\end{eqnarray}
Here the ratio $L/\Lambda$ is a constant which can be fixed
only by experiment.

\bigskip
{\bf 3.Discussion.} Following \cite{MTY} let us form the
difference between the radial pressure
$\tau=-T_{\hat{\rho}\hat{\rho}}^{\rm ren}$ and energy density
$\varrho=T_{\hat{t}\hat{t}}^{\rm ren}$ and integrate it from
$\rho=\rho_0<0$ to $\rho=+\infty$. The resulting value turns out
to be definitely positive as the sum of the two positive terms:

\begin{equation}
\int_{\rho_0}^{\infty}{(\tau-\varrho)\exp{(-\Phi)}d\rho}=
\left.-{5\over 72\pi^2}{r^{\prime}\over r^3}\exp{(-\Phi)}
\right |_{\rho=\rho_0}+{4\over 3}\int_{\rho_0}^{\infty}
{{r^{\prime 2}\over r^4}\exp{(-\Phi)}d\rho} > 0.
\end{equation}
That is, averaged weak energy condition for the electromagnetic
vacuum in the wormhole topology is violated as well as the local
weak energy condition at the throat is \cite{Kh}.

Thus, a necessary condition for the electromagnetic vacuum to be
able to support the wormhole geometry is fullfilled. This
requirement is stronger one than that of violation of the local
weak energy condition {\it anywhere} (however, the fact that the
local weak energy condition is violated {\it namely at the
throat} does not follow, generally speaking, from violation of
the averaged weak energy condition). Remarkably that it is
fullfilled irrespectively of the detailed form of the metric
$r(\rho)$, $\Phi (\rho)$ at least in the considered orders of
quasiclassical expansion.

Moreover, the Einstein equations with $T_{\mu\nu}^{\rm ren}$
(\ref{T-ren}) as the source formally have the wormhole-type
solution with $r={\rm const}$ (infinitely long wormhole).
However, the derivatives $\Phi^{(n)}$ are not small for this
solution, and approximation based on the few first orders in
the expansion over derivatives is questionable. Evidently, any
progress in this direction is connected with finding more
accurate and universal expression for the stress-energy tensor.

\bigskip
I am grateful to Professors V.L.Chernyak and I.B.Khriplovich for
the interest to the work and the discussion.

\end{document}